\documentstyle[epsfig]{aipproc}

\begin{document}
\title{Jet Production in DIS at NLO
\footnote{Talk given by D.Z. at the {\it 5th International Workshop on Deep 
Inelastic Scattering and QCD}, Chicago, Illinois, 14--18 April 1997.}
}
\author{Erwin Mirkes$^1$ and Dieter Zeppenfeld$^2$}
\address{
$^1$ Institut f\"ur Theor. Teilchenphysik, 
     Universit\"at Karlsruhe, D-76128 Karlsruhe, Germany \\
$^2$ Physics Department, University of Wisconsin, Madison, WI 53706, USA}

%\lefthead{LEFT head}
%\righthead{RIGHT head}
\maketitle

\begin{abstract}
Dijet production in DIS is an important laboratory for testing our 
understanding of perturbative QCD. Flexible NLO Monte Carlo programs allow
to investigate general jet definition schemes. For forward jet production
at $p_{Tj}\approx Q$ and $x_{jet}\gg x$ the NLO predictions fall well below
HERA data, thus providing evidence for BFKL dynamics.
\end{abstract}

%\section*{Introduction}
Multi-jet production in DIS is an important topic of study
at HERA. Good event statistics allow for precise measurements
and thus for a variety of tests of our understanding of QCD dynamics. 
Such tests include 
a) the determination of $\alpha_s(\mu_R)$ from dijet production 
over a range of renormalization scales~\cite{alphas},
b) the measurement of the 
gluon density in the proton (via $\gamma g\to q\bar q$)~\cite{mikunas},
c) forward jet production in the low-$x$ 
regime as a signal of BFKL dynamics~\cite{H1-forward,woelfle},
d) the determination of $\alpha_s(\mu_R)$ and power corrections in
DIS event shapes \cite{rabbertz}.
For quantitative studies one clearly needs to compare data
with calculations which include next-to-leading order (NLO) QCD 
corrections.

Here we concentrate on issues related to dijet 
production at HERA. The present status of NLO Monte Carlo programs is 
reviewed. Second, we compare predictions in different jet definition
and parton recombination schemes. Finally, the ${\cal O}(\alpha_s^2)$
predictions for forward jet production are discussed.

\vspace*{-0.1in}

\section*{NLO Monte Carlo programs}

\vspace*{-0.1in}

NLO QCD corrections to one and two-jet production cross sections
and distributions, for $\gamma^*$ exchange, are implemented in 
the fully differential $ep \rightarrow n$ jets event generators
MEPJET~\cite{mepjet} and 
DISENT~\cite{disent}. Full neutral current exchange ($\gamma^\ast$ 
and/or $Z$) is now available in MEPJET 2.0 for tree level cross sections 
up to ${\cal O}(\alpha_s^3)$, {\it i.e.} up to 4-jet final states. 
NLO corrections to two-jet production including $Z$ and $W$ exchange 
will be available soon. Both programs allow to study arbitrary 
experimental cuts and jet definition schemes. Previous 
calculations~\cite{previous} were limited to a JADE type 
algorithm and important single jet-mass effects were neglected.
Because of the disagreement with these earlier calculations a comparison
of the two new calculations is particularly important. 

In Table~\ref{table_comp} we compare results from MEPJET and DISENT 
for 2- and 3-jet rates at HERA ($E_e=26.7$~GeV, $E_p=820$~GeV) in 
the $k_T$-scheme with $y_{cut}=1$ and a hard scattering scale 
$E_T^2=Q^2$. MRS D$_{-}^{'}$ parton distributions are used~\cite{mrs}, 
we fix $\alpha_{QED}=1/137$, and require $Q^2>40$~GeV$^2$. 
Factorization and renormalization scales are set to $\mu^2_F=\mu^2_R=Q^2$.
No further cuts are imposed in the first column while in the second column
reconstructed jets must have transverse momenta $p_{Tj}>5$~GeV
and pseudo-rapidities $|\eta_j|<3.5$. We find agreement at the 2--3\%
level, which is slightly larger than, but still compatible with the 
statistical errors determined by the Monte Carlo integration routines.
For all practical applications this agreement is satisfactory. 
\begin{table}[t]
\begin{tabular}{l|cc|cc}
\hline
     &   &   &  \mbox{$p_{Tj}>5$~GeV} &\mbox{$|\eta_j|<3.5$} \\
     &  \mbox{DISENT}  & \mbox{MEPJET} &\mbox{DISENT}  & \mbox{MEPJET} \\
\hline\\
\mbox{LO 2 jet}
    & 395.6 $\pm$ 1.5 pb & 392.6 $\pm$ 0.5 pb
    & 339.0 $\pm$ 0.4 pb & 337.5 $\pm$ 0.4 pb       \\
\mbox{LO 3 jet}
    & 33.0 $\pm$ 0.5 pb & 32.5 $\pm$ 0.2 pb 
    & 27.7 $\pm$ 0.2 pb & 27.2 $\pm$ 0.2 pb  \\
\mbox{NLO 2 jet incl.}
    & 561 $\pm$ 7 pb  &  559 $\pm$ 6 pb
    & 463 $\pm$ 8 pb & \mbox{480 $\pm$ 5 pb}       \\
\hline
\end{tabular}
\caption{Comparison of DISENT~\protect\cite{disent} and 
MEPJET~\protect\cite{mepjet} predictions for $n$-jet cross sections 
in DIS at HERA. See text for details. }
\label{table_comp}
\end{table}

\vspace*{-0.1in}

\section*{Jet Definition Schemes}
\vspace*{-0.1in}

The internal structure of jets is first modeled at NLO where two massless
partons may be recombined to form a jet. The dependence of jet cross sections
on the details of the internal jet structure can be investigated by studying
different recombination schemes: the $E$-scheme where parton 4-momenta 
are added to form massive jet 4-momenta, and the $E0$ and $P$-schemes in 
which the jets are made massless by rescaling either the 3-momentum or the
energy of the resulting cluster. 

We use MEPJET for numerical studies with MRS D-' parton distribution 
functions~\cite{mrs}. The renormalization and factorization scales are 
set to one half the sum of parton transverse momenta in the Breit frame.
A  minimal set of kinematical cuts is imposed.
We require 40~GeV$^2<Q^2<2500$ GeV$^2$,
$0.04 < y < 1$, an energy cut of $E(l^\prime)>10$~GeV on the scattered 
lepton, and pseudo-rapidities $|\eta|<3.5$  %$\eta=-\ln\tan(\theta/2)$
for the scattered lepton and jets. Also, jets 
must have transverse momenta of at least 2~GeV in the lab and the 
Breit frame.

Within these general cuts four different jet definition schemes are 
considered.
i) A cone algorithm with a jet separation cut of $\Delta R<1$
and a minimum jet transverse momentum of $p_{Tj}>5$~GeV in the lab 
frame. 
ii) The $k_T$ algorithm (in the Breit frame) as defined in Ref.~\cite{kt} 
with a hard scattering scale $E_T^2=40$~GeV$^2$
and a resolution parameter $y_{cut}=1$ for resolving the macro-jets. 
In addition, jets are required to have a minimal 
transverse momentum of 5 GeV in the Breit frame.
iii) The $W$-scheme where parton/cluster pairs (including the 
proton remnant) with invariant mass 
squared, $M_{ij}^2=(p_i+p_j)^2 < y_{cut}W^2$ are
recombined~\cite{previous}. 
iv) The ``JADE'' algorithm \cite{jade} which is
obtained from the $W$-scheme by replacing the invariant
definition $M_{ij}^2$ by $2E_iE_j(1-\cos\theta_{ij})$. In the $W$ and JADE
schemes we set $y_{cut}=0.02$.
\begin{table}[t]
\begin{tabular}{l|ccccc}
     &  \mbox{\phantom{spac}2-jet\phantom{sp} }
     &  \mbox{\phantom{sp}2-jet excl.\phantom{sp}}
     &  \mbox{\phantom{sp}2-jet incl.\phantom{sp}}
     &  \mbox{\phantom{sp}2-jet incl.\phantom{sp}} 
     &  \mbox{\phantom{sp}2-jet incl.\phantom{sp}} \\
     &  \mbox{LO}
     &  \mbox{NLO} ($E$)
     &  \mbox{NLO} ($E$)
     &  \mbox{NLO} ($E0$)
     &  \mbox{NLO} ($P$)\\
\hline\\
\mbox{cone} & 1107~pb & 1047~pb & 1203~pb & 1232 pb  & 1208 pb \\
$k_T$       & 1067 pb & 946 pb  & 1038 pb & 1014 pb  & 944 pb \\
$W$         & 1020 pb & 2061 pb & 2082 pb & 1438 pb  & 1315 pb \\
\mbox{JADE} & 1020 pb & 1473 pb & 1507 pb & 1387 pb  & 1265 pb \\
\hline
\end{tabular}
\caption{Two-jet cross sections in DIS at HERA. Results are given at LO and
NLO for the four jet definition schemes and acceptance cuts described in 
the text. The 2-jet inclusive cross section at NLO is given for three 
different recombination schemes.
}\label{table_dijet}
\end{table}

The resulting cross sections (see Table~\ref{table_dijet}) show large NLO
corrections for the JADE and $W$-schemes while $K$-factors close to 
unity are found for the $k_T$ and cone algorithms. Another 
disadvantage of the $W$ and JADE schemes is their strong recombination
scheme dependence. Since even in a NLO calculation the internal structure of
a jet is only calculated at tree level, these strong variations point to large
uncertainties from two-loop effects in the $W$ and JADE schemes. In both 
schemes widely separated but relatively soft partons tend to be clustered
to what is then defined as a jet, even though these clusters can be very 
massive and quite distinct from the pencil-like and low-mass objects which
one starts out with in the parton model or at LO~\cite{mepjet}. 
These differences then result in the large higher order corrections to jet 
cross sections shown in Table~\ref{table_dijet}. These effects are much 
smaller in the cone and the $k_T$ schemes which are hence favored for 
precision QCD studies in DIS. 
\vspace*{-0.1in}

\section*{Probing BFKL in Forward Jet Production}
\vspace*{-0.1in}

Recently much interest has been focused on the small
Bjorken-$x$ region, where one would like to distinguish 
BFKL evolution~\cite{bfkl}, which resums the leading $\alpha_s \ln 1/x$
terms, from the more traditional DGLAP evolution equation~\cite{dglap},
which resums leading $\alpha_s \ln Q^2$ terms.
BFKL evolution can be enhanced and DGLAP evolution suppressed by studying 
DIS events which contain an identified jet of longitudinal momentum 
fraction $x_{jet}=p_z(jet)/E_{proton}$ (in the proton direction) 
which is large compared to Bjorken $x$~\cite{mueller}. When
tagging a forward jet with $p_{Tj}\approx Q$ this leaves little room for
DGLAP evolution while the condition $x_{jet}\gg x$ leaves BFKL evolution 
active. This leads to an enhancement of the forward jet production cross 
section proportional to $(x_{jet}/x)^{\alpha_P -1}$ over the DGLAP 
expectation.

\begin{figure}[t]
\epsfxsize=4.0in
\epsfysize=3.0in
\begin{center}
\hspace*{0in}
\epsffile{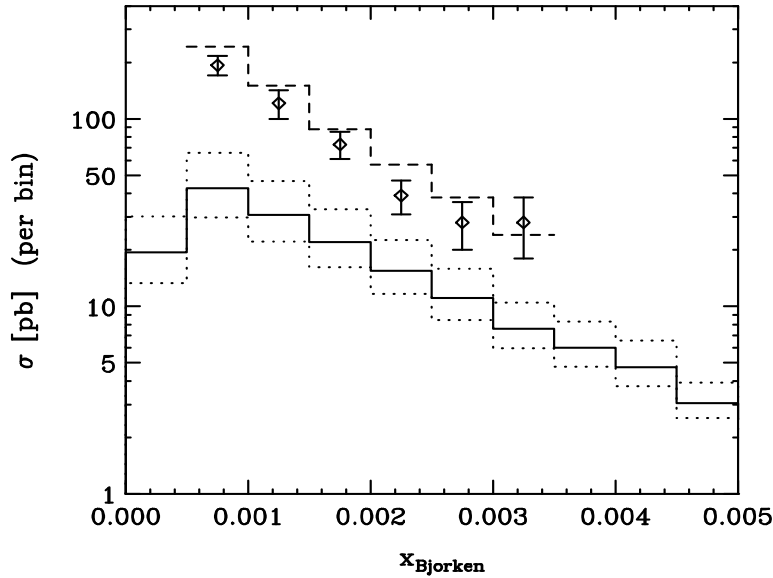}
\vspace*{0.5cm}
\caption{
Forward jet cross section at HERA as a function of Bjorken $x$ within the H1 
acceptance cuts~\protect\cite{H1-forward}. 
The solid histogram gives the NLO MEPJET result for  
the scale choice $\mu_R^2=\mu_F^2=\xi(0.5\sum k_T)^2$ with $\xi=1$. 
The two dotted histograms show the uncertainty of the NLO prediction, 
corresponding
to a variation of $\xi$ between 0.1 and 10. The BFKL result of 
Bartels et al.~\protect\cite{bartelsH1} is shown as the dashed 
histogram. The data points are the new, preliminary 
H1 measurements~\protect\cite{H1-forward}.
\label{fig:h1comp}
}
\vspace*{-0.1in}
\end{center}
\end{figure}

A conventional fixed order QCD calculation up to ${\cal O}(\alpha_s^2)$ 
does not yet contain any BFKL resummation and must be 
considered a background for its detection; one must search
for an enhancement in the forward jet production cross section above the 
expectation for two- and three-parton final states. 

The full calculation of the forward jet inclusive
cross section in DIS, at ${\cal O}(\alpha_s^2)$, has been performed in
Ref.~\cite{MZ-prl}. 
Ordinarily, such a calculation would contain 3-parton final states at tree 
level, 1-loop corrections to 2-parton final states and 2-loop corrections to
1-parton final states. However, these 2-loop contributions vanish identically,
once the condition $x\ll x_{jet}$ is imposed. The remaining 2-parton and 
3-parton differential cross sections,  and the cancellation of
divergences between them, are the same as those entering 
a calculation of 2-jet inclusive rates. These elements are already 
implemented in the MEPJET program, which, therefore, can be used to determine 
the inclusive forward jet cross section at ${\cal O}(\alpha_s^2)$.

In Fig.~\ref{fig:h1comp} numerical results are compared with recent data 
from H1~\cite{H1-forward}. Here the conditions $p_{Tj}\approx Q$ and 
$x_{jet}\gg x$ are satisfied by selecting events with forward jets (in the 
angular region $7^o < \theta_j < 20^o$) with
\begin{eqnarray}
   0.5     & < & p_{Tj}^2/Q^2\; < \; 2\;, \label{eq:fjb} \\
   x_{jet} & \approx & E_j/E_{proton} > 0.035\;. \label{eq:fja}
\end{eqnarray}
Clearly H1 observes substantially more forward jet events than expected 
from NLO QCD. However, since H1 accepted jets of rather low transverse 
momentum, $p_{Tj}>3.5$~GeV, the comparison of our parton level calculation 
to the data may be subject to sizable hadronization corrections.
A recent BFKL calculation~\cite{bartelsH1} 
(dashed histogram) agrees better with the data, but here the overall 
normalization is uncertain and the agreement may be fortuitous. 
A very rough estimate of the uncertainty of the NLO calculation is 
provided by the two dotted lines which correspond to variations 
by a factor 10 of the renormalization and factorization scales
$\mu_R^2$ and $\mu_F^2$. 

Very similar results are found by ZEUS~\cite{woelfle}.
We conclude that the HERA data show evidence for BFKL dynamics in 
forward jet events via an enhancement in the observed forward jet cross 
section above NLO expectations. However, additional data, with harder
forward jets, are needed to make the comparison of QCD calculations with
data more reliable.

\vspace*{-0.1in}

\section*{Acknowledgments} 
\vspace*{-0.1in}
We thank M.~Seymour for his help comparing MEPJET and DISENT. 
This research was supported by the University of Wisconsin 
Research Committee with funds granted by the Wisconsin Alumni Research 
Foundation and by the U.~S.~Department of Energy under Grant 
No.~DE-FG02-95ER40896. The work of E.~M. was supported in part  
by DFG Contract Ku 502/5-1.

\end{document}